\documentclass[aps, prl, reprint, authoryear, showpacs, showkeys, 10pt, floatfix, nofootinbib]{revtex4-2}

\usepackage[english]{babel}
\usepackage[utf8x]{inputenc}
\usepackage[T1]{fontenc}

\usepackage{graphicx}
\usepackage{dsfont}
\usepackage{epstopdf}
\usepackage{indentfirst}
\usepackage{amsmath}
\usepackage{amsfonts}
\usepackage{amssymb}

\usepackage{soul}
\soulregister\cite7
\soulregister\ref7
\soulregister\pageref7
\usepackage[colorlinks=true, allcolors=blue]{hyperref}

\usepackage{color}
\usepackage[dvipsnames]{xcolor}

\usepackage{xargs}            \usepackage{booktabs}

\def\gamlq{\gamma_{\ell}(q)}

\def\Gamq{\Gamma_q}

\def\RSIN#1#2#3{ \frac{\sin\left(#1\, #2\, #3 \right) }{  \sin\left(
#2\, #3 \right)} }

\usepackage[normalem]{ulem}

\begin{document}

\title{
Basin sizes depend on stable eigenvalues in the Kuramoto model
}

\date{\today}

\author{Antonio Mihara$^1$} 
\author{Michael Zaks$^2$}
\author{Elbert Macau$^{3}$}
\author{Rene O. Medrano-T$^{1,4}$}
\affiliation{$^1$Departamento de Física, Universidade Federal de São Paulo,UNIFESP, 09913-030, Campus Diadema, São Paulo, Brasil}
\affiliation{$^2$Institute of Physics, Humboldt University of Berlin, 12489 Berlin, Germany}
\affiliation{$^3$Universidade Federal de São Paulo, UNIFESP, 12247-014, Campus São José dos Campos, São Paulo, Brazil}
\affiliation{$^4$Departamento de Física, Instituto de Geociências e Ciências Exatas,
Universidade Estadual Paulista, UNESP, 13506-900, Campus Rio Claro, São Paulo, Brasil}
%\email[Author e-mails: ]{mihara74@gmail.com, rene.medrano@unifesp.br}

\begin{abstract}

We show
that for the Kuramoto model (with identical phase oscillators equally coupled) its global statistics and size of the basins of attraction can be estimated through the
eigenvalues of all stable (frequency) synchronized states.
This result is somehow unexpected since, by doing that, one could just
use local analysis
to obtain global dynamic properties. But recent works based on Koopman and Perron-Frobenius operators
demonstrate that global features of a nonlinear dynamical system, with some
specific conditions, are somehow encoded in the local eigenvalues of its
equilibrium states.
Recognized numerical simulations in the literature reinforce our analytical results.

\end{abstract}

\maketitle

\section*{Introduction}

Since the pioneering studies of Winfree on biological rhythms
in the end of 60's \cite{Winfree1967}, phase oscillators have been playing a central role in the study of collective behavior. The most famous derivation, the Kuramoto model \cite{Kuramoto1975,Kuramoto1984} and its variants, has been successfully employed to understand problems related to synchronization in various areas
of science. It includes synchronization in flashing fireflies \cite{Ermentrout1991}, circadian rhythms \cite{Antonsen2008}, swarming dynamics \cite{Keeffe2017}, cardiac pacemaker cells \cite{Osaka2017}, superconducting Josephson junctions \cite{Wiesenfeld1998}, power-grid networks \cite{Dorfler2013}, and the Millenium bridge oscillation \cite{Strogatz2005}. In front of this vast diversity of these dynamical systems, emerges a relevant question that guides this paper: what are the conditions that lead each system to the correct operation?

Basin of attraction, the set of initial conditions %\red{
from which the solutions %} 
converge asymptotically to a %\red{
given %} 
attractor, is an intricate and fundamental concept in dynamics. Although the definition is straightforward, the boundaries of the basin as well as its measure, may be difficult %} 
to study even in low-dimensional systems %\red{
and also for such simple attractors as stable equilibrium states. %}. \red{
Since the basin can include the points quite distant from the attracting set, the size of the basin, as a general rule, is not determined by the local properties of the attractor. %}. 
In dissipative maps and flows, it is delimited by the complex geometrical configuration of stable
manifolds of unstable invariant sets, %generally resulting 
which may imply in fractal boundaries \cite{Ott2002}. This feature makes statistics a proper method to evaluate quantities in a basin of attraction. That approach has been applied in the Kuramoto model of coupled phase oscillators, in the context of stable synchronized twisted states \cite{Wiley2006, Delabays2017,Ochab2010}. These studies focused on the size of the basin, lately interpreted in networks as the basin stability: the likelihood quantification of returning to the same synchronized state \cite{Menck2013}. 

In particular, in 2006 Wiley, Strogatz, and Girvan \cite{Wiley2006} investigated the Kuramoto model of $N$ identical phase oscillators on a ring, each one equally coupled with the $R$ nearest neighbors on either side [Eq. (\ref{eq:WSG})]. The authors studied, through numerical experiments and also analytically, for different low values of $R$ some relevant aspects of the so-called sync basin (the attraction basin of the state of full synchronization  
%\red{
for which, in the appropriate co-rotating reference frame, all oscillators share the steady phase value). %with $\theta_{i}=C, \forall\,i$) 
They showed that for $R/N>0.34$, the sync basin is the whole phase space, except for a set of measure zero. Below this critical value the stable %\red{
steady configurations %} 
called $q$-twisted states, characterized by a constant %$\theta_{j} - \theta_{j-1} \neq 0$)  \red{
difference of phases between the neighboring units, %}
emerge in the phase space. %\red{
The number of twists $q$ counts overall rotations around the circle that occur while passing along the ensemble from the first to the last unit. %}
The simulations revealed that ({\it i}) the probability of the final twisted state having $q$ twists follows a Gaussian distribution $\sim\exp[-q^2/(2\sigma^2)]$ with respect to the winding number $q$, and ({\it ii}) the standard deviation $\sigma$ of this distribution scales as $\sqrt{N/R}$, namely $\sigma \sim 0.2 \, \sqrt{N/R}$. %\red{
Remarkably, this finding was supported by 
%had no more than 
a heuristic argument for such statistical patterns, 
leaving rigorous derivations of ({\it i}) and ({\it ii}) as open questions. %}

For the %\red{
next-neighbor %} 
coupling ($R=1$), this problem was revisited in \cite{Delabays2017} with an accurate numerical method to measure the volume of %\red{
the basin for %} 
each stable state. The authors obtained a typical linear size
[$\alpha_\tau(q)$] for each basin of attraction of $q$-twisted stable state, so that the volume of the basin of attraction of each stable state is proportional to $V_q \sim \alpha_\tau^N(q)$. 
Besides these studies another important step in the 
knowledge of the basin of attraction of the Kuramoto model, still locally coupled, was given by Ochab and Góra \cite{Ochab2010}. They observed 
%\sout{that there is} 
a direct correlation between the size of basins of attraction and the respective eigenvalues of %\red{
the Jacobian matrices of stable %} 
$q$-twisted states: solutions having the maximal negative eigenvalue 
%of the perturbation matrix 
closer to zero (\textit{less stable}) feature smaller basins of attraction than the \textit{more stable} solutions for which all eigenvalues are strongly negative. 
In general one does not expect that local properties of equilibrium states have direct relations with global properties of the state space, but that result is consistent with recent mathematical results about global
properties of certain nonlinear systems on compact manifolds which shall be discussed further and will also serve as basis for the study that we present here.   
Keeping in mind that in the Kuramoto model the only dynamic action is the attractive 
or repulsive interaction between the nodes,  and
that the most attractive configuration possesses 
the largest size basin of attraction, we delve into this idea and suggest a theoretical description for the size of the basin of attraction in the Kuramoto model.

Following we start with a brief summary of the Kuramoto model
and of recent mathematical results concerning global and local 
properties of certain nonlinear systems on compact manifolds.
Using some approximations, we obtain an analytical expression that has many similarities with the basin volume distribution obtained by 
\cite{Wiley2006}, mentioned above.
Then our analytical results are compared with the numerical experiments  
strengthening the evidence for the strong correlation between eigenvalues and 
basin sizes and providing more
arguments towards an explanation of open questions
({\it i}) and ({\it ii}).
Our approach
should work with systems that can be reduced to the Kuramoto model, as done recently in an experimental network of nanoelectromechanical oscillators \cite{Matheny2019}, but we expect that our approach can be applied to other systems, as discussed in Final Remarks.

\section{Theoretical aspects}

\subsection*{Kuramoto model: solutions and eigenvalues}
 
%\red{
Following %} 
\cite{Wiley2006}, we consider here a system of $N$ identical Kuramoto oscillators
 on a regular ring 
 where each oscillator is coupled with equal strength to its $R$ nearest neighbors on either side. %\red{
 In the co-rotating reference frame %} 
 the time evolution of this system is governed by the following set of ODEs
\begin{equation}
    \dot{\theta_j}  = %\frac{g}{N-1} 
    \sum_{k=j-R}^{j+R} G \,\sin ( \theta_k - \theta_j ) \, ,
    \,\, j=1,2,...,N ;
    \label{eq:WSG}
\end{equation}
where the index $k$ is periodic mod $N$, %\red{
the coupling constant %} 
$G$  %\red{
is positive 
%}. 
%\blue{\footnotesize [
and can be completely removed from the equations by rescaling the time. %Should we?]}.

As pointed out in~\cite{Wiley2006}, the set of
equations (\ref{eq:WSG}) is a gradient system that can be recast as 
$\dot{\theta} = - \nabla V$ %\red{
with e.g. $V(\theta_1,\ldots,\theta_N)=G\sum_{i,j}\cos(\theta_i-\theta_j)$ so that $V$ is bounded both from below and from above: $-N^2G\leq V\leq N^2G$. %}.\red{
Therefore $dV/dt=-(\nabla V)^2$ %} 
and all trajectories, %\red{
except the points of equilibrium
and their stable manifolds,  
are flowing ``monotonically downhill'' 
%\sout{on the potential $V$ surface}
and asymptotically %\red{
tend to those of the equilibria that correspond to the local minima of $V$. %}.\red{
For this reason, %}
``we need not concern ourselves with the
possibility of more complicated long-term behavior, such as
limit cycles, attracting tori, or strange attractors'' for 
(\ref{eq:WSG}) \cite{Wiley2006}.

The system (\ref{eq:WSG}) has a %\red{
family %} 
of equilibrium states  %(in the rotating frame) 
which can be characterized by an integer $q$
\begin{equation}
    \theta_j = \frac{2\pi q}{N} j + C \, , 
\label{eq:solu}
\end{equation}
the ``winding number'', which measures the number of full twists in phase as one goes around the ring once and can assume the values $q=0,1,2,..., N-1$; and $C$ is a real constant.
The state with $q=0$ corresponds to
$\theta_j = C, \forall j$, i.e.\ with all oscillators 
synchronized in phase. 

In the states with $q\neq 0$ all oscillators are synchronized in frequency but with a constant
phase difference between two successive units:
$\Delta_j \equiv \theta_j - \theta_{j-1} = 2\pi q/N ,
\forall\, j$. Such states are also known as ``$q$-twisted'' 
states. Recalling that $\Delta_j$ is taken mod $2\pi$, we notice that for a state with $q = N-n, \, 1 \leq n \leq N/2$, the phase
difference between two successive oscillators is
\begin{equation*}
 \Delta_j = \frac{2\pi}{N} (N-n) = 2\pi - \frac{2\pi}{N} n
 = - \frac{2\pi}{N} n < 0 \, ,
\end{equation*}
%\red{
so that %} 
the phases of oscillators are distributed (in the order:
$\theta_1, \theta_2, ...$) clockwise around 
the circle, in such a way that the winding number of
this state can also be represented by $q=-n$, 
denoting $n$ clockwise twists. 
So one can change the range of winding numbers to
$q  = -m,..., -1, 0, 1,... , m$ with $ m = N/2$
(for odd $N$, $m=(N-1)/2$).

In turn, the eigenvalues of %\red{
the Jacobian matrices near %} 
those equilibrium states 
(parametrized by $q$) are real and %\red{
obey the expression %}
~\cite{Mihara2019} 
\begin{eqnarray}
\gamma_{\ell} (q) = -4\, G\, \,\sum_{k=1}^R 
\cos\left(k\frac{2\pi}{N}q\right)\sin^2\left(k \frac{\pi}{N}\ell\right)  \, , \nonumber\\
\ell = 1,2,..., N-1 .
\label{eq:gamma}
\end{eqnarray}

Clearly, the linear stability of these equilibria is 
determined by the eigenvalues, and it was
observed that the stability of each $q$-twisted state depends 
on the ratio $R/N$ ($\propto$ connectivity) : (i) for 
$R/N \gtrsim 0.34$, $q=0$ (all oscillators in phase) is 
the only stable equilibrium state; (ii) below this threshold, 
and as the ratio decreases, more twisted states (with $q\neq 0$) 
become stable
\cite{Wiley2006,Maistrenko2012,Mihara2019}.
As mentioned before, the authors of  Ref.~\cite{Wiley2006} 
concluded that the distribution of volumes of basins of 
attraction of the $q$-twisted states has
Gaussian shape
with standard deviation $\sigma \sim 0.2 \sqrt{N/R}$.

\subsection*{The stability measure}

The Koopman (or composition) operator approach can provide a global 
description of dynamical systems in
terms of the time evolution of observables
(functions) of the state space. In this approach a nonlinear
dynamical system is represented in terms of an infinite--dimensional
(but linear) operator acting on a Hilbert space
of functions of the system states. The spectral decomposition of the
Koopman operator provides complete description of the nonlinear system.
Despite being an infinite--dimensional operator,
there are several numerical methods 
(such as DMD and its variations) capable
of obtaining finite--dimensional approximations for Koopman 
eigenvalues / modes
and they have applications in various real-world problems such as fluids 
dynamics, power grids, epidemiology, climatology, etc. ,
see Ref.\cite{koopmanbook2020} and references therein. 
In turn the Perron-Frobenius (or transfer) operator evolves 
densities of trajectories in the state space. It is also linear 
and is dual to the Koopman operator.
Both operators share the same spectral properties and can provide
global descriptions of a dynamical system \cite{koopmanbook2020}.

Recently it was demonstrated that for a Morse-Smale gradient flow
acting on a smooth, compact and oriented manifold with no boundary,
the spectrum of the transfer operator is given by linear combinations
of the Lyapunov exponents at the critical points of the Morse function
(i.e., the eigenvalues of Jacobian at the fixed points) 
\cite{DangRiviere2019} and it holds globally on the manifold. 
This result agrees with the observation
that for a $d$-dimensional autonomous system with a hyperbolic fixed point $x^* \in X$,
where $X$ is a compact, connected and forward--invariant subset of 
$\mathds{R}^d$, the spectrum of the Koopman operator is given by the
eigenvalues of the Jacobian matrix evaluated at $x^*$ \cite{mauroy2016}
and also with a relevant property of this operator: if
$\phi_1$ and $\phi_2$ are Koopman eigenfunctions associated with the
eigenvalues $\mu_1$ and $\mu_2$, then $\Phi =\phi_1^a \phi_2^b$, 
with $a,b \in \mathds{R}$, 
is also a Koopman eigenfunction with eigenvalue $a\mu_1 + b \mu_2$.

On the other hand one should notice that the set of Kuramoto equations 
(\ref{eq:WSG}) above is a Morse-Smale system
with Morse function given by the ``potential'' $V(\theta_1,...,\theta_N)$,
the critical points of $V$ are the $q$--twisted states and the 
Lyapunov exponents (at critical points) are the eigenvalues $\gamma_{\ell}(q)$. 
Then the mathematical results \cite{DangRiviere2019,mauroy2016} above 
ensure that the $\gamma_{\ell}(q)$, despite being obtained by local methods, 
somehow contain global information of system (\ref{eq:WSG}), and we shall use 
them to explore the basins of attraction of the $q$-twisted states.
In order to represent the stability in all directions of the phase space, 
we consider the sum of the
eigenvalues of a $q$-twisted state,
\begin{equation}
\hat{\gamma}_q \equiv \sum_{\ell=1}^{N-1}\, \gamlq
\end{equation}
which resembles the entropy functional for
Morse-Smale diffeomorphisms in the framework of (Gibbs) 
variational principle for dynamical systems \cite{takahashi84}.
\begin{figure}[b]
\centering
   \includegraphics[width=0.8\linewidth]{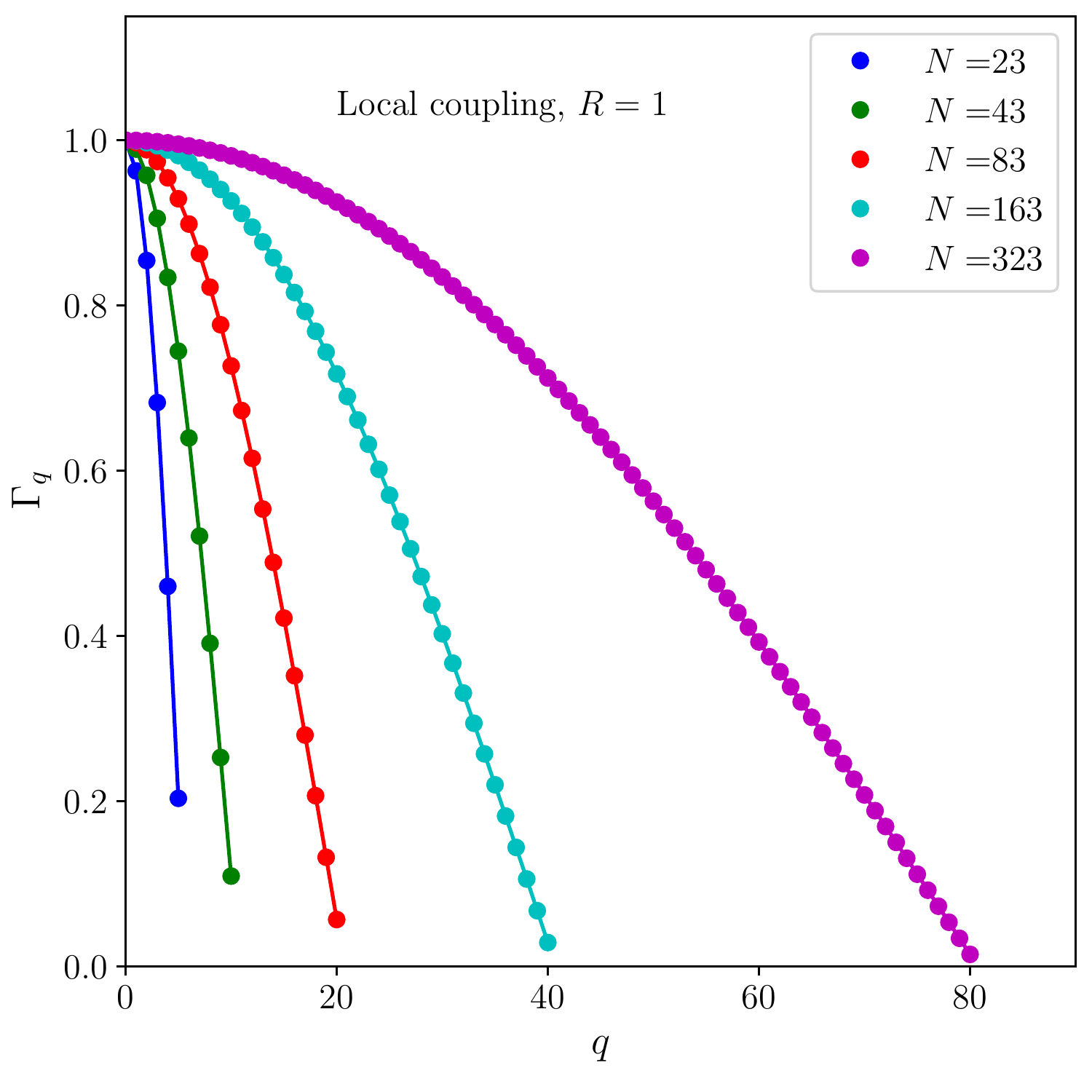}
   \includegraphics[width=0.8\linewidth]{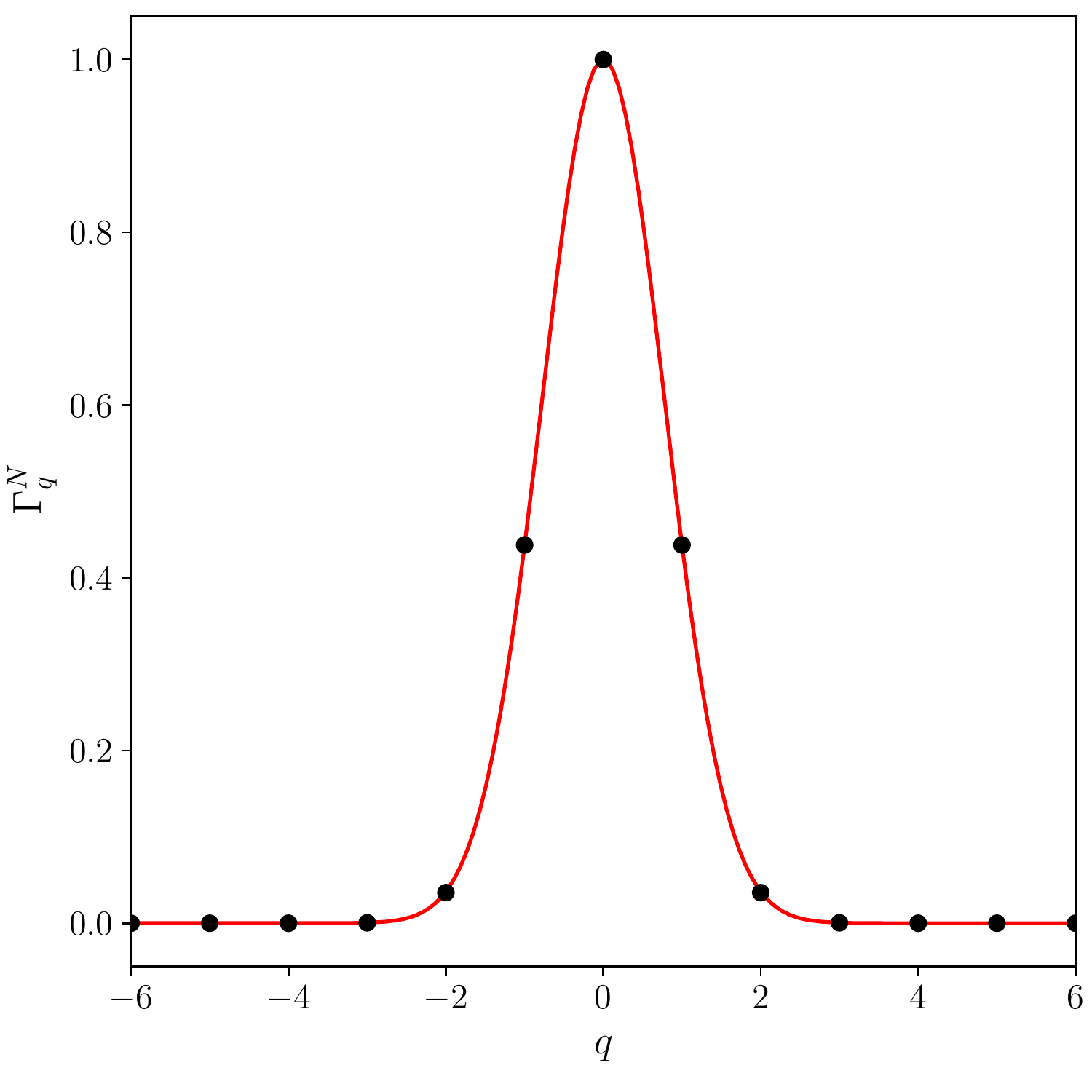}
\caption{(a) Dependence of $\Gamma_q$ with the winding number $q$, for different network sizes. By comparing this plot
with that in Fig.3 (inset) of Ref.~\cite{Delabays2017}, one 
can notice that $\Gamma_q$ behaves as the linear size of
the basin of attraction of the $q$ states.
(b) Plot of $\Gamma_q^N$, for $N=60$ and $R=2$ (black circles), and the fitted Gaussian curve (red line).}
\label{fig:Gammaq}
\end{figure}
%%%%%%%%%%%%%%%%%%%%%%%%

 Then we
 define $\Gamma_q$, the {\it equilibrium stability measure}, as the sum of the eigenvalues of a $q$-twisted state, normalized by the most negative $\hat{\gamma}_q$ which in this case is $\hat{\gamma}_0$:
\begin{equation}
\Gamma_q \equiv \frac{\hat{\gamma}_q}{\hat{\gamma}_0} \, .
\label{eq:Gammaq}
\end{equation}

We present in Fig.~\ref{fig:Gammaq}(a) the plot of $\Gamma_q$ with respect to $q$ for different network sizes with local coupling
($R=1$).
(hyperbolic)  stable states, i.e. those with 
$\gamma_{\ell}<0, \forall\,\ell$, were taken into account.
It is remarkable the similarity between the plot 
of $\Gamma_q$ and the plot of $\alpha_{\tau}(q)$, the typical linear size of the basin of attraction presented in Fig.~3 (inset) of 
Ref.~\cite{Delabays2017}.   
Therefore, if $\Gamma_q$ has a behavior similar to the linear size of
a basin of attraction, it is reasonable to expect that for
a network with $N$ oscillators the volume of the 
basin of attraction of a given $q$-state will behave as $\sim \Gamma_q^N$. In Fig.~\ref{fig:Gammaq}(b) we plot 
$\Gamma_q^N$, for $N=60$ and $R=2$; that can be well approximated
by a Gaussian curve. To establish this result, in the next subsection 
we show explicitly  for low values of $R/N$ that
$\Gamma_q^N$  
can be approximated by a Gaussian function with respect to the winding number $q$, 
.

\subsection*{The Gaussian distribution of the basins of attraction}

Our goal here is to derive an explicit expression for the standard deviation $\sigma$ in terms of the number of nodes $N$ and the connection $R$ of the network. 
Considering $G = 1/N$, Eq. (\ref{eq:gamma}) returns
\begin{equation}
\gamlq = -4\, \sum_{k=1}^R \, \frac{1}{N}\, 
\cos( k\, q\, \delta )\, \sin^2 (k\, \ell\,\delta/2 ) \, ,
\label{eigenv1}
\end{equation}
where $\ell = 1, 2, \dots, N-1$ and $\delta = 2\pi/N$.

By using %\sout{some} 
trigonometric identities, Eq. (\ref{eigenv1}) can be rewritten as
\begin{align}
\gamlq =& - \frac{1}{N}\, \sum_{k=1}^R \, \{
2\cos( k\, q\, \delta )\, - \cos( k\,\delta\, (q+\ell) ) \nonumber \\
&- \cos( k\,\delta\, (q-\ell) )
\} \, , 
\end{align}
and on performing the summations one obtains
\begin{align}
\gamlq =& - \frac{1}{N}\, \left\{
\RSIN{M}{q}{\pi/N} - \frac{1}{2}\left[
\RSIN{M}{(q+\ell)}{\pi/N}\right.\right. \nonumber \\
&\left.\left.+\RSIN{M}{(q-\ell)}{\pi/N}\right]
\right\} \, ,
\label{eigenv2}
\end{align}
where $M \equiv 2R+1$.

Now let us %\red{
evaluate %} 
$\hat{\gamma}_q = \sum_\ell \gamma_\ell(q)$. The summation of the first term in the RHS of Eq. (\ref{eigenv2}) is trivial, 
but for the second and the third terms, the sums are approximated by 
integrals since we are regarding $N \gg R,\, q$ :
\begin{align}
B_{\pm} =& \sum_{\ell=1}^{N-1}\, \frac{1}{N}\, \RSIN{M}{(q\pm\ell)}{\pi/N} \nonumber \\
\approx& \int_0^{1-\varepsilon} \RSIN{M}{(y \pm x)}{\pi} dx \, ,
\end{align}
where $\varepsilon = 1/N$, $x = \ell/N$, $y = q/N$ and $M=2R+1$.
Then one arrives at
\begin{equation}
\hat{\gamma}_q = 1 - \RSIN{M}{y}{\pi} + \mathcal{O}(\varepsilon^2) \, .
\end{equation}

In turn, the normalized sum of eigenvalues $\Gamq$, can be 
expanded around $y \sim 0$
\begin{eqnarray}
\Gamq &=& 1 - \frac{\pi^2}{3} (2R+1)(R+1)\, y^2 + \dots \,  \nonumber\\
&\approx& 1 - \beta \frac{q^2}{N} \, ,  
\end{eqnarray}
where
\begin{equation}
\beta  = \frac{\pi^2}{3} \frac{(2R+1)(R+1)}{N}\, .
\end{equation}

For $N \gg R, q$, the quantity $\Gamq^N$ can be approximated
by a Gaussian function
\begin{equation}
\Gamq^N \approx \left( 1 - \beta \frac{q^2}{N} \right)^N 
\approx  e^{\displaystyle -\beta q^2} \, ,
\end{equation}
with standard deviation equal to $ 1/{\sqrt{2\beta}}$. 
As a result, %Then our theoretical result for 
the standard deviation can be
written as
\begin{equation}
\sigma_T = \sqrt{N}\, \mathcal{F}(R),\quad
\mathcal{F}(R) = 
\left[\frac{3}{2\pi^2(2R+1)(R+1)} \right]^{1/2}.
\label{sigma}
\end{equation}

\section{Comparison with {available numerical} data}
\label{sec:exp}

To compare this result [Eq.\ (\ref{sigma})] with the one %\red{
known from %} 
the literature \cite{Wiley2006}, we restrict ourselves to small values of $R$ 
and obtain the following expansion for $\sigma_T$
\begin{eqnarray}
\sigma_T &\approx& \frac{4\sqrt{3}}{1369\pi} \left( 125\frac{1}{\sqrt{R}} - 26 
\right) \sqrt{N} \nonumber\\
&\approx& 0.2014 \sqrt{ \frac{N}{R} } - 0.04188 \sqrt{N} \, ,
\label{sigmalin}
\end{eqnarray}
which has approximately the same scaling law ($\sigma \sim 0.2\sqrt{N/R}$) of the volume of the basin of attraction obtained through numerical experiments. 
It is important to remark that those experiments were carried out with datasets from different network sizes. 
But, as we can see, the second term of Eq.(\ref{sigmalin}) is almost negligible and does not differ much for small values of $N$. 
Therefore, $\sigma$ can be interpreted as linearly dependent 
%\red{
on $\sqrt{N/R}$ as argued in \cite{Wiley2006}. 
Nevertheless, Eq.(\ref{sigmalin}) indicates that actually $\sigma/\sqrt{N}$ increases (almost) linearly with $1/\sqrt{R}$. 

\begin{figure}[ht]
\centering
   \includegraphics[width=0.9\linewidth]{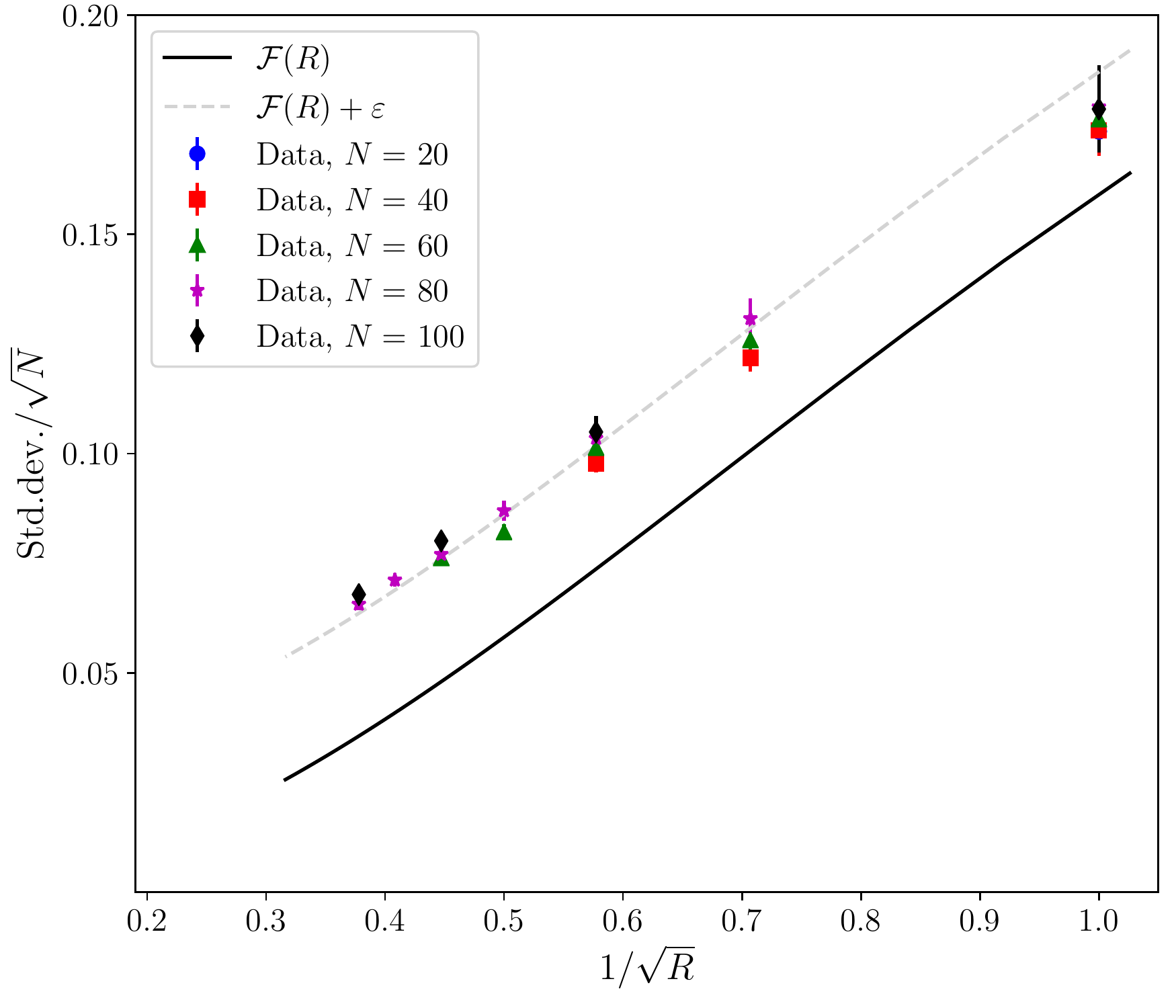} \\
\caption{Standard deviation of the distribution of
volumes of basins of attraction, divided by 
$\sqrt{N}$. The blue, red, green, magenta and black marks refer, respectively, to numerical experiments performed in networks with $N= 20, 40, 60, 80,$ and 100 with $R/N \lesssim 0.1$.
The solid black line is the theoretical result given by Eq.~(\ref{sigma}). Gray dashed line: a good agreement between 
our results and experimental data could be obtained by adding a constant $\varepsilon \approx 0.028$ to $\mathcal{F}(R)$.}
\label{fig2}
\end{figure}

In Fig.\ref{fig2} we can observe a qualitative agreement 
(same scaling $\sigma \sim 0.2\sqrt{N/R}$) between the theoretical standard deviation Eq.(\ref{sigma}) and  numerical experiments for different network sizes, with $R/N \lesssim 0.1$. %\red{
Furthermore, %}
a good agreement between theory and experimental data
can be obtained by adding a constant $\varepsilon$ ($\approx 0.028$) to $\mathcal{F}(R)$ in Eq.(\ref{sigma}), as illustrated by the gray dashed
line in Fig.\ref{fig2}. Then, based on the eigenvalues
of equilibrium states, a good estimate for the
standard deviation of the distribution of volumes
of basins of attraction can be written as
\begin{eqnarray}
\frac{\sigma}{\sqrt{N}} &=& \mathcal{F}(R) \, + \,\varepsilon \nonumber\\
&=& \left[\frac{3}{2\pi^2(2R+1)(R+1)} \right]^{1/2}
 \, + \,\varepsilon \, .
\label{eq:epsilon}
\end{eqnarray}

\section{Final Remarks}
\label{sec:final}

Based on recent mathematical results that establish relationships between global and local properties
of nonlinear flows on compact manifolds \cite{DangRiviere2019,mauroy2016} and
also on the observation
that the size of the basin of attraction of a $q$--twisted state is correlated with its eigenvalues \cite{Ochab2010}, we define the stability measure $\Gamma_q$ as 
proportional to the sum of eigenvalues of the $q$ states and
observe that $\Gamma_q$ (our Fig.~\ref{fig:Gammaq}(a)) behaves similarly to the linear basin size $\alpha_{\tau}(q)$ 
(see inset of Fig.~3 in Ref.~\cite{Delabays2017}).
Then for small $q, R \ll N$, we found an analytic expression for $\Gamma_q^N$, a Gaussian distribution for $q$ with standard deviation
that scales as $\sim 0.2 \sqrt{N/R}$, the same behavior obtained
by numerical simulations in Ref.~\cite{Wiley2006}.
In Fig.~\ref{fig2} we showed that a good agreement between
our theoretical result and experimental data could be obtained 
by adding a small constant $\varepsilon$, Eq.~(\ref{eq:epsilon}).
 This indicates that $\Gamma_q^N$ (obtained from the equilibria eigenvalues) is successful in capturing how the basin volumes are distributed according to the size $N$ and the topology $R$ of the network.

A priori it is not expected that global dynamical  properties can be obtained from local %\red{
characteristics, %} 
such as %\red{
the Jacobian %} 
eigenvalues. But our results indicate that some global properties of the system (\ref{eq:WSG}) are somehow reflected/encoded in the 
(local) eigenvalues of the equilibria and is compatible with results about
global dynamics of Morse-Smale gradient systems \cite{DangRiviere2019}.
On the other hand,
 we suspect that some facts contribute to $\Gamma_q^N$ having many characteristics identical to the distribution of basins of attraction of the $q$-twisted states. The phase space of the system (\ref{eq:WSG}) is
\begin{itemize}
    \item a compact manifold: %\red{
          an $N$-torus, %},
    \item ``smooth'' because, as pointed out in
    Ref.~\cite{Wiley2006}, the system (\ref{eq:WSG}) %\red{
    features %} 
    gradient dynamics with trajectories flowing monotonically over a potential surface and asymptotically reaching fixed points, %\red{
    both in forward and backward time. %}.
    No complicated behavior (limit cycles, attracting tori, strange attractors, etc.) occurs.
    \item most likely the basins are well-defined regions in the phase space, separated by smooth high-dimensional hypersurfaces: %\red{
    segments of codimension-1 stable manifolds of the equilibria that possess just one positive Jacobian eigenvalue. These segments are matched on codimension-2 stable manifolds of the equilibria with two positive eigenvalues, and so on. Moreover, in high-dimensional convex bodies the bulk of the volume lies in the immediate vicinity of the boundaries: for a 40-dimensional sphere with radius $R$ or a cube with the size $2R$ the thin boundary layer $0.9R<|r|<R$ contains over 98\% of the volume. Therefore, regions of the phase space adjacent to the basin boundaries are responsible for the dominating part of the basin volume. Such geometry favors %} 
    the long distance linear behavior. We do not expect these results to replay in systems with complicated basins of attraction delimited by fractal, riddled, or Wada boundaries.
    
\end{itemize}

Our approach has revealed that %\red{
certain %} 
global phenomena in networks of phase oscillators can be understood by local studies  where the dynamics is strongly dominated by attractive and repulsive interactions between the nodes. 
In such systems the action of the eigenvalues predicting the dynamics %\red{
protrudes over %} 
large distances from the equilibrium state, whether an attractor, a saddle or a repeller, as shown in \cite{Mihara2019}. 
Therefore, we hope that this study can be extended to other 
network systems with similar features (high-dimensional, with phase space that is a compact manifold, etc.) mainly for the case of Morse-Smale gradient systems
with a finite number of hyperbolic equilibrium states.

%%%

%%%%%%%%%%%%%%%%%%%%%%%%%%%%%%%%%%%%%%%%%%%%%%%%%%%%%%%%%%%%%%%%%%%%%

\section*{Acknowledgements}

ROM-T.\ acknowledges the support by S\~ao Paulo Research Foundation 
(FAPESP, Proc. 2015/50122-0) and thanks the support and the fruitful period from 09/21 to 10/15 of 2019 at the Physics Department of the Humboldt University of Berlin where part of this research was developed. EENM acknowledges the support of Sao Paulo Research Foundation (FAPESP, Proc 2018/03517-8) and CNPq. AM thanks FAPESP for financial support 
during the workshop V ComplexNet (from 
August 26th to September 1st, 2018, 
in Cachoeira Paulista-SP) where part 
of this work was carried out.
The authors also thank Prof. Michael Rosenblum for important discussions.
The plots were created with Python and its libraries: Matplotlib, Numpy and Scipy.

%%%
% \bibliography{sample.bib}

%%%%%

\end{document}